\documentclass[doublecol]{epl2}

\usepackage{latexsym,amsmath,amssymb}

\title{Spin diode behavior in transport through single-molecule magnets}
\shorttitle{Spin diode behavior in transport through SMMs} 

\author{M. Misiorny\inst{1} \and I. Weymann\inst{1,2} \and J. Barna\'{s}\inst{1,3}}
\shortauthor{M. Misiorny \etal}

\institute{
  \inst{1} Faculty of Physics, Adam Mickiewicz University, 61-614 Pozna\'{n}, Poland\\
  \inst{2} Physics Department, Arnold Sommerfeld Center for Theoretical Physics and Center for NanoScience,
           Ludwig-Maximilians-Universit\"{a}t M\"{u}nchen, 80333 M\"{u}nchen, Germany\\
  \inst{3} Institute of Molecular Physics, Polish Academy of Sciences, 60-179 Pozna\'{n}, Poland
}

\abstract{ We study transport properties of a single-molecule
magnet (SMM) weakly coupled to one nonmagnetic and one
ferromagnetic lead. Using the diagrammatic technique in real time,
we calculate transport in the sequential and cotunneling regimes
for both ferromagnetic and antiferromagnetic exchange coupling
between the molecule's LUMO level and the core spin. We show that
the current flowing through the system is asymmetric with respect
to the bias reversal, being strongly suppressed for particular
bias polarizations. Thus, the considered system presents a
prototype of a SMM spin diode. In addition, we also show that the
functionality of such a device can be tuned by changing
the position of the molecule's LUMO level and strongly depends on
the type of exchange interaction.
}%

\begin{document}

\maketitle

\section{Introduction}
\begin{figure}[t]
    \onefigure[width=0.75\columnwidth]{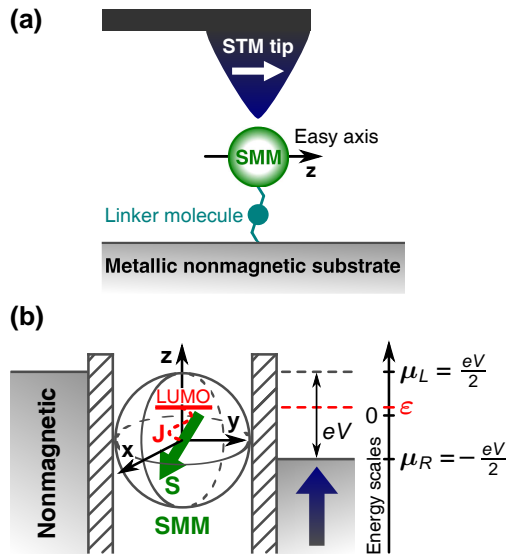}
    \caption{(Color online) (a) Illustration showing a possible realization of
    a SMM-based device employing one metallic magnetic lead: the SMM molecule
    attached to a metallic \emph{nonmagnetic} surface is pinned from the top
     by a \emph{magnetic} STM tip. For simplicity, the magnetization of the
     tip and the molecule's easy axis are assumed here to be parallel.
     (b) Schematic of a SMM attached to one nonmagnetic (left)
    and one ferromagnetic (right) lead.
    The current flows due to tunneling through the LUMO level
    which is exchange coupled to the molecule's core spin $S$.}
    \label{Fig:1}
\end{figure}

Due to their particular physical properties, such as an energy
barrier for the spin reversal or long relaxation
times~\cite{Gatteschi_book}, single-molecule magnets (SMMs) are
inherently predestined for applications in novel molecular
electronic or spintronic circuits~\cite{Bogani_NatureMater7/08,Mannini_NatureMater8/09}.
Up to now, several different physical mechanisms employing SMMs as
a key component have been theoretically considered. It has been in
particular shown that the SMM's spin can be reversed by means of
spin polarized
currents~\cite{Timm_PRB73/06,Misiorny_PRB75/07,Misiorny_PSSb246/09},
or by applying a spin bias~\cite{Lu_PRB79/09}. Furthermore, when
bridged between two nonmagnetic metallic leads a SMM can work as a
spin filter~\cite{Barrazalopez_JAP105/09,Barrazalopez_PRL102/09}.
Recent experiments on electronic transport through SMMs
connected to  metallic but nonmagnetic
leads~\cite{Heersche_PRL96/06,Ni_APL89/06,Jo_NanoLett6/06,Henderson_JApplPhys101/07,Voss_PRB78/08}
clearly show that the aforementioned ideas are in principle
experimentally feasible. From a conceptual point of view, the
simplest realization of a system consisting of a SMM attached to
one ferromagnetic and one nonmagnetic reservoir could be a device
involving the scanning tunneling microscope (STM) with a magnetic
tip and a SMM on a metallic but nonmagnetic substrate,
Fig.~\ref{Fig:1}(a). The advantage of such a geometry is that by
choosing an appropriate ligand shell for the molecule, one can
obtain the specific orientation (e.g. parallel) of the molecule's
easy axis with respect to the
surface~\cite{Fonin_Polyhedron28/09}. Furthermore, there are
experimental techniques which allow for deposition of a film of
well-dispersed SMMs on a substrate, enabling access to individual
molecules with the STM
tip~\cite{Zobbi_ChemCommun12/05,Abdi_JAP95/04,Cornia_StructBond122/05,Naitabdi_AdvMater17/05,Burgert_JAmChemSoc129/07}.
Thus, it is interesting to consider how transport properties of a
SMM change if one of two metallic nonmagnetic leads is replaced by
a magnetic one.

In this Letter we study in general
transport through a SMM weakly coupled to electrodes with unequal
spin polarizations. As already shown in the case of quantum
dots~\cite{Rudzinski_PRB64/01,Wilczynski_JMMM290/05,Souza_PRB75/07,Hamaya_PRL102/09},
transport properties of such systems exhibit a significant
asymmetry with respect to the bias reversal. In addition, due to
coupling to ferromagnetic leads and the spin dependence of
tunneling processes, the current flowing through such a diode
becomes spin polarized and, interestingly, the spin polarization
may change with reversing the bias voltage. In fact, very recently
spin diode behavior was predicted and observed experimentally in
another class of molecular structures, namely in single-wall
carbon nanotubes~\cite{Merchant_PRL100/08,Weymann_APL92/08}. In
this work we propose a {\it molecular} spin diode based on single
molecule magnets. The considered system consists in particular of
a SMM weakly coupled to one nonmagnetic and one ferromagnetic lead
of high spin polarization; Fig.~\ref{Fig:1}(b). To determine the
transport properties of such a system we employ the real-time
diagrammatic technique (RTDT), which allows us to systematically
take into account the sequential and cotunneling processes
contributing to current. The sequential tunneling dominates the
current for voltages larger than the Coulomb correlation energy
and becomes exponentially suppressed in the Coulomb blockade
regime, where transport is mainly due to cotunneling processes. As
presented in the following, unequal couplings to the leads give
rise to a pronounced asymmetry of spin-polarized current with
respect to the bias reversal.
Furthermore, in addition to the analysis of system's $I$-$V$
transport characteristics, we also discuss the zero-frequency
noise of tunneling current associated with discreteness of charge
carriers (shot noise). The shot noise proves to be a source of
useful information about various transport properties of the
system, such as effective charges, coupling strengths
or various types of correlations, which is hardly attainable from
direct measurement of electric
current~\cite{Blanter_PhysRep336/00}.

\section{Model and theoretical method}

We assume that transport through the molecule occurs only
\emph{via} the lowest unoccupied molecular orbital (LUMO) level,
described by the energy $\varepsilon$ and the local spin operator
$\textbf{s}=\frac{1}{2}\sum_{\sigma\sigma'}c_\sigma^\dag
\bm{\sigma}_{\sigma\sigma'} c_{\sigma'}$, which is coupled to the
molecule's internal spin $\textbf{S}$ \emph{via} exchange
interaction $J$. In principle, the exchange interaction can be
either ferromagnetic or antiferromagnetic, depending on the
specific type of a SMM. The molecule is thus fully characterized
by Hamiltonian
\begin{align}\label{eq:Hamiltonian_SMM}
    \mathcal{H}_\textrm{SMM} =& -\Big[D +\sum_{\sigma}D_{1}\, c_\sigma^\dag
    c_\sigma + D_{2}\, c_\uparrow^\dag c_\uparrow c_\downarrow^\dag c_\downarrow
    \Big]S_z^2
    \nonumber\\
    &+ \sum_{\sigma} \varepsilon\,c_\sigma^\dag c_\sigma
    + U\, c_\uparrow^\dag c_\uparrow c_\downarrow^\dag c_\downarrow
    -J \textbf{s}\cdot\textbf{S},
\end{align}
where $D$ is the uniaxial anisotropy constant of a neutral
molecule, $D_1$ and $D_2$ stand for corrections to the anisotropy
due to single and double occupation of the LUMO level,
respectively, while $U$ describes the Coulomb correlations between
two electrons occupying the LUMO level. We note that although the corrections $D_1$ and $D_2$
 to the anisotropy constant $D$ are not
 crucial in observing the presented effects, we take
 them into account to make the model realistic.
In Eq.~(\ref{eq:Hamiltonian_SMM}) we neglected small transverse
anisotropy~\cite{Misiorny_PSSb246/09}. Furthermore, it is assumed
that the easy axis of the SMM is collinear with the magnetization
of the right electrode, see Fig.~\ref{Fig:1}(b), and both electrodes
are described by noninteracting electrons, $
\mathcal{H}_\textrm{el}=\sum_q\sum_{\textbf{k}, \sigma}
\varepsilon_{\textbf{k}\sigma}^q\: a_{\textbf{k}\sigma}^{q\dag}
a_{\textbf{k}\sigma}^q$, with $\varepsilon_{\textbf{k}\sigma}^q$
being the energy of an electron with the wave vector $\textbf{k}$
and spin $\sigma$ belonging to the lead $q$. Finally, the
Hamiltonian describing tunneling processes between the molecule
and the leads reads, $
\mathcal{H}_\textrm{T}=\sum_q\sum_{\textbf{k},\sigma} \big[T_q\,
a_{\textbf{k}\sigma}^{q\dag}c_\sigma^{} + T_{q}^* c_\sigma^\dag
a_{\textbf{k}\sigma}^q\big]$, where $T_q$ denotes the tunnel
matrix elements between the molecule and the $q$th lead. Coupling
between the molecule and each lead can be expressed by
$\Gamma_\sigma^q=2\pi|T_q|^2\rho_\sigma^q$ and
$\Gamma_{q}=(\Gamma^{q}_{+}+\Gamma^{q}_{-})/2$. The spin
polarization of the right magnetic lead is defined as
$p=(\rho_+^R-\rho_-^R)/(\rho_+^R+\rho_-^R)$, where $\rho_{+(-)}^R$
is the density of states for the majority (minority) electrons.

To calculate the current flowing through the system we employ the
real-time diagrammatic
technique~\cite{Schoeller_PRB50/94,Koenig_PRB54/96}. This approach
allows us to take into account the first and second order
tunneling processes through SMM in a fully systematic
way~\cite{Misiorny_PRB79/09}. Within the RTDT, the occupation
probabilities $P_\chi$ of finding the system in a many-body state
$|\chi\rangle$ are given by,
\begin{equation}
    (\mathbf{\tilde{\Sigma}}\mathbf{P})_{\chi} =
    \Gamma\delta_{\chi\chi_0},
\end{equation}
where $\mathbf{\tilde{\Sigma}}$ is
the self-energy matrix accounting for various tunneling processes
in the system, with a row $\chi_0$ replaced by
$(\Gamma,\dots,\Gamma)$ due to normalization of probabilities. The
current can be then calculated as
\begin{equation}
    I=-\frac{ie}{2\hbar}\,{\rm Tr}\{\mathbf{\Sigma}^{\rm I}\mathbf{P}\},
\end{equation}
with $\mathbf{\Sigma}^{\rm I}$ being the modified self-energy to
take into account the number of electrons transferred through the
system. Also the expression for the current noise in the limit of
low frequencies~\cite{Blanter_PhysRep336/00},
\begin{equation}
    S=2\int_{-\infty}^0\textrm{d}t\big[\big\langle I(t)I(0)+I(0)I(t)\big\rangle-2\langle I\rangle^2\big]
\end{equation}
can be conveniently reformulated in the language of RTDT, for
details see Ref.~\cite{Thielmann_PRB68/03}. At this point, we
ought to note that in the next section, rather than the shot noise
$S$, we present the deviation of the current noise from its
Poissonian value, described by the Fano factor  $F=S/2e|I|$.

Finally, to determine the current order by order in tunneling
processes, one performs perturbation expansion of the
self-energies and probabilities in the coupling strength
$\Gamma$~\cite{Schoeller_PRB50/94,Koenig_PRB54/96}. Here, we have
taken into account the first and second order terms of expansion,
which correspond to sequential and cotunneling processes,
respectively~\cite{Misiorny_PRB79/09,Weymann_PRB78/08}.

\section{Results and discussion}

Since the model under consideration applies to various types of
SMMs, here we present results only for the conceptually easiest
case, though capturing essential physics, i.e. for a molecule with
$S=2$ and strong uniaxial magnetic anisotropy. Moreover, to keep
the discussion most intuitive, we neglect the role of electron
charge sign, assuming $e>0$.

\begin{figure}[t]
  \onefigure[width=0.94\columnwidth]{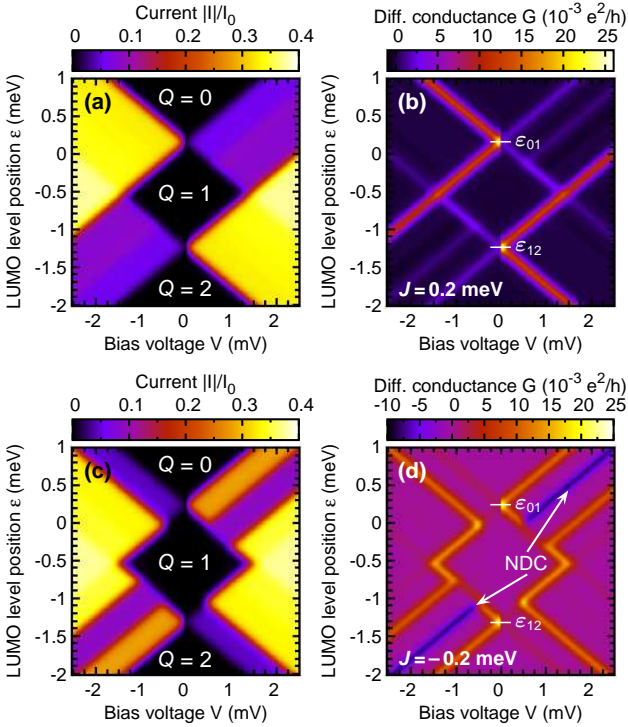}
  \caption{
  (Color online) (a,c) The absolute value of the current $I$
  in units of $I_0=2e\Gamma/\hbar\approx 0.5$ nA, and (b,d) differential conductance $G$
  as a function of the bias voltage
  $V$ and the LUMO level position $\varepsilon$ for (a)-(b) \emph{ferromagnetic} ($J>0$)
  and (c)-(d) \emph{antiferromagnetic} ($J<0$) exchange coupling, $|J|=0.2$ meV.
  $Q$ in (a,c) represents the average charge accumulated in the LUMO level.
  The other parameters are: $S=2$,  $D=50$ $\mu$eV, $D_1=-5$ $\mu$eV, $D_2=2$ $\mu$eV,
  $U=1$ meV, $k_{\rm B}T=40$ $\mu$eV, $p=0.9$, and $\Gamma=\Gamma_L=\Gamma_R=1$ $\mu$eV.
  \label{Fig:2}}
\end{figure}

\begin{figure}[t]
  \onefigure[width=0.94\columnwidth]{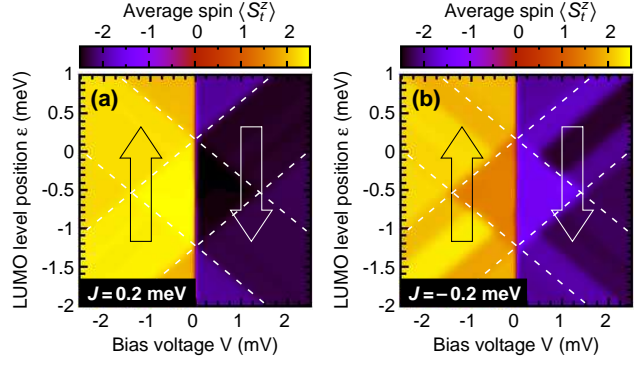}
  \caption{
  (Color online) The $z$th component of the SMM's total spin
  $S_t^z=S_z+(c_\uparrow^\dagger c_\uparrow^{}-c_\downarrow^\dagger c_\downarrow^{})/2$
  as a function of the bias voltage $V$ and the LUMO level position $\varepsilon$ for
  (a) \emph{ferromagnetic} and (b) \emph{antiferromagnetic} exchange coupling.
  Hollow arrows depict the preferred direction of the SMM's spin with respect
  to the right lead's magnetic moment, whereas the dashed lines correspond to the position
  of main peaks in the differential conductance, Fig.~\ref{Fig:2}(b,d) -- i.e. they divide
  regions representing different occupation $Q$ of the LUMO level.
  The remaining parameters are as in Fig.~\ref{Fig:2}.}
  \label{Fig:3}
\end{figure}

First of all, we note that due to large spin asymmetry in coupling
of the SMM to the ferromagnetic lead, Fig.~\ref{Fig:1}(b), the
tunneling probability for spin-majority (spin-up) electrons is
much larger than that for spin-minority (spin-down) electrons. On
the other hand, the rate of tunneling processes between the
molecule and the nonmagnetic lead is the same for both spin
orientations. This generally leads to an asymmetry of tunneling
current with respect to the bias reversal.

\begin{figure*}[t]
  \onefigure[width=0.65\textwidth]{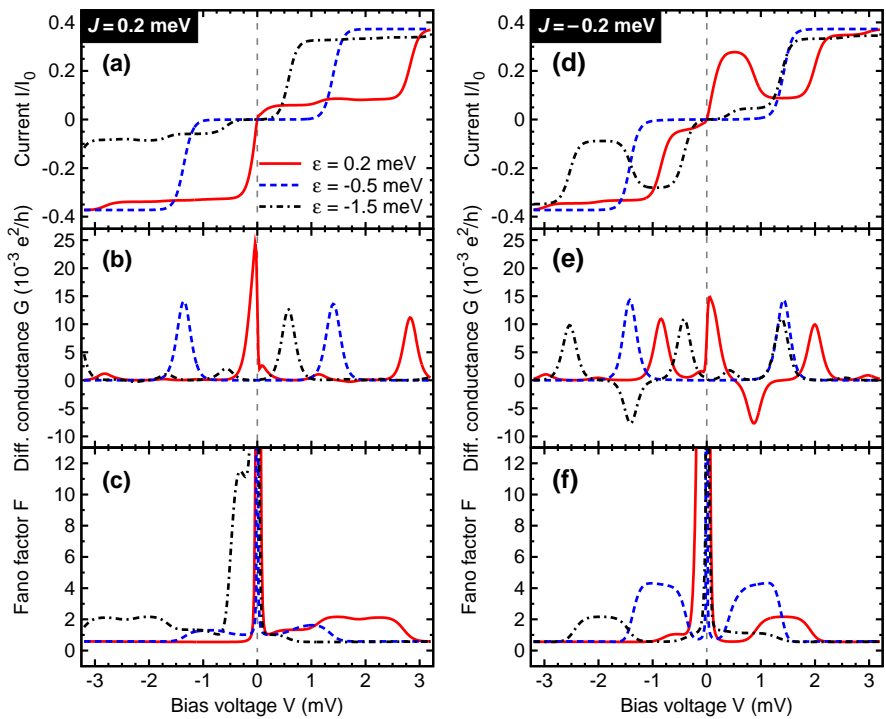}
  \caption{
  (Color online) Selected cross-sections of density plots in Fig.~\ref{Fig:2}
  for specific values of the LUMO level position $\varepsilon$,
  which depict the current $I$ flowing through the system (a,d),
  differential conductance $G$ (b,e), and Fano factor $F$ (c,f)
  as a function of the bias voltage $V$ for the \emph{ferromagnetic}
  (left panel) and \emph{antiferromagnetic} (right panel) exchange couplings.
  The other parameters are the same as in Fig.~\ref{Fig:2}.
  At low bias voltages the shot noise is dominated by thermal fluctuations
  while the current vanishes leading to a divergency in the Fano factor when $V\to 0$.}
  \label{Fig:4}
\end{figure*}

In Fig.~\ref{Fig:2}(a)-(b) we display the current and differential
conductance as a function of the LUMO level position and bias
voltage for ferromagnetic ($J>0$) interaction between the LUMO
level and the SMM's spin. Close to the first resonance,
$\varepsilon\approx\varepsilon_{01}$, see Fig.~\ref{Fig:2}(b),
current can flow easily from the ferromagnetic lead to the
nonmagnetic one ($V<0$), while it is suppressed for the opposite
direction ($V>0$); see also solid line in Fig.~\ref{Fig:4}(a). To
understand this behavior one should take into account the
following facts. First, the molecule's states with one extra
electron in the LUMO level correspond  to the total spin number
$S=5/2$ and $S=3/2$, with the former being of lower energy.
Second, orientation of the molecule's spin depends significantly
on the current direction, Fig.~\ref{Fig:3}(a), and for  $V>0$ the
SMM's spin tends towards antiparallel orientation with respect to
the electrode's magnetic moment, whereas for $V<0$ the spin
prefers the parallel alignment. Thus, for low bias transport can
occur mainly via the state corresponding to $S=5/2$, and this
requires spin-down electrons for $V>0$ and spin-up electrons for
$V<0$. Unfortunately, tunneling rate for spin-down electrons is
significantly reduced due to fewer available states in the
minority spin band of the ferromagnetic lead, which effectively
leads to the suppression of current for $V>0$. When bias voltage
increases, then the state corresponding to $S=3/2$ becomes active
in transport as well, leading only to a small increase of the
current due to the spin selection rules for tunneling
processes. However, the blockade for positive bias is removed
when double occupancy of the LUMO level is admitted, which takes
place for bias voltages exceeding some threshold value.
Furthermore, the electron flow from ferromagnetic lead to the
nonmagnetic one is spin polarized and degree of this polarization
depends mainly on the spin polarization of the lead. In fact, in
the case of a perfect halfmetallic ferromagnet ($p\to 1$), the
current flowing towards halfmetallic lead would be totally
blocked. Moreover, different time scales associated with
spin-majority and spin-minority electrons lead in turn to
considerable current fluctuations and super-Poissonian shot noise,
see Fig.~\ref{Fig:4}(c) for $\varepsilon = 0.2$ meV.

The situation becomes significantly different when the LUMO level
is doubly occupied in equilibrium; $\varepsilon=-1.5$ meV in
Fig.~\ref{Fig:4}(a). Now, the behavior of current is reversed as
compared to the case of $Q=0$, since the current is suppressed for
$V<0$, i.e. for electrons tunneling from the magnetic lead. This
is associated with the fact that an electron first has to tunnel
out of the LUMO level and then another electron can enter the
molecule. Thus, for positive bias a spin-up electron can easily
tunnel out to the ferromagnetic lead. On the other hand, when the
bias is reversed and the spin-down electron tunnels out of the
molecule leaving it in the state corresponding to $S=5/2$, the
current becomes suppressed, as the rate for tunneling of spin-down
electrons from the ferromagnetic lead to the molecule is
relatively small. This also leads to super-Poissonian shot noise,
as shown in Fig.~\ref{Fig:4}(c).

More complex transport characteristics are observed when the
exchange interaction is antiferromagnetic ($J<0$);
Fig.~\ref{Fig:2}(c)-(d). The most striking difference is the
appearance of additional peaks in the current when the LUMO level
is initially either empty or doubly occupied, Figs.~\ref{Fig:2}(c)
and~\ref{Fig:4}(d), which are accompanied by negative differential
conductance (NDC); Figs.~\ref{Fig:2}(d) and \ref{Fig:4}(e).
Consider first the case of empty LUMO level in equilibrium,
$\varepsilon=0.2$ meV in Fig.~\ref{Fig:4}(d). The key difference
is that now the molecule's state  corresponding  to the total spin
number $S=3/2$ has lower energy and determines transport
properties at low voltages. Thus spin-up electrons are involved in
charge transport for $V>0$ and spin-down electrons for $V<0$.
Consequently, the current is suppressed for negative bias and can
easily flow for positive one. When the bias voltage reaches values
admitting transport through the $S=5/2$ state, the current for
positive bias becomes suppressed by a spin-down electron tunneling
to the LUMO level, while suppression for negative voltage becomes
then lifted. In turn, when bias increases further admitting doubly
occupation of the LUMO level, the blockade for positive bias
becomes removed as well. Transport characteristics for  doubly
occupied LUMO level in equilibrium can be explained in a similar
way.

On the other hand, when the LUMO level is singly occupied and its
position corresponds approximately to the middle of the Coulomb
blockade [dashed line in Figs.~\ref{Fig:4}(a,d)], the diode
behavior disappears and the current recovers symmetry with respect
to the bias reversal regardless of the type of the exchange
interaction $J$. This is due to the fact that now with increasing
the bias voltage all charge states of the molecule start taking
part in transport at the same time, i.e. once the bias voltage
reaches the threshold. The transport characteristics become then
symmetric with respect to the bias reversal and the noise is
rather sub-Poissonian, indicating the role of single-electron
charging effects in transport. Note, however, that in the Coulomb
blockade regime bunching of inelastic cotunneling processes may
still result in enhancement of the shot noise; see dashed lines in
Figs.~\ref{Fig:4}(c,f). Furthermore, it is visible that in the
case of $\varepsilon=-0.5$ meV the enhancement is much more
pronounced in the case of the antiferromagnetic coupling,
Fig.~\ref{Fig:4}(f). Such a behavior stems from the fact that for
$J<0$ both energetically lowest lying molecular magnetic states
$S_t^z=\pm3/2$ allow the possibility of occupying the LUMO level
by an electron either with the spin up or down, whereas for $J>0$
the state $S_t^z=+5/2$ ($S_t^z=-5/2$) can only accommodate an
electron with spin up (down). For this reason, in the former case
both majority and minority electrons of the ferromagnetic lead can
participate in transport, thus increasing the fluctuations.

To conclude, we have proposed a {\it molecular} spin diode based
on single molecular magnets coupled to one ferromagnetic and one
nonmagnetic leads. We have shown that the spin-dependent current
flowing through the device becomes suppressed for one bias
polarization and is enhanced for the opposite one. The suppressed
transport is then accompanied with super-Poissonian shot noise.
Moreover, we have demonstrated that the transport characteristics
of SMM spin diodes strongly depend both on the number of electrons
occupying the LUMO level, and on the type of exchange interaction
between the LUMO level and the SMM's core spin. This gives the
possibility to tune the functionality of a SMM spin diode by
shifting the position of the LUMO level with a gate voltage.

Finally, we would like to note that although the presented results
were calculated for a specific molecule with $S=2$, the spin diode
behavior can be observed in a large class of molecules. The main
requirement is to have considerable spin asymmetry in the
couplings to the left and right electrodes and well-defined spin
states in the molecule, i.e. $J,D \gg T$, where $T$ is the
experimental temperature.

\acknowledgments
This work, as part of the European Science Foundation EUROCORES
Programme SPINTRA, was supported by funds from the Ministry of
Science and Higher Education as a research project in years
2006-2009 and the EC Sixth Framework Programme, under Contract N.
ERAS-CT-2003-980409. The authors also acknowledge support from the
Adam Mickiewicz University Foundation (M.M.), funds from the
Ministry of Science and Higher Education as research projects in
years 2008-2009 (M.M.) and  2008-2010 (I.W.), and the Foundation for
Polish Science (I.W.).


\begin{thebibliography}{10}
\expandafter\ifx\csname url\endcsname\relax\def\url#1{\texttt{#1}}\fi

\bibitem{Gatteschi_book}
\Name{Gatteschi D., Sessoli R. \and Villain J.} \Book{{Molecular Nanomagnets}}
  (Oxford University Press, New York) 2006.

\bibitem{Bogani_NatureMater7/08}
\Name{Bogani L. \and Wernsdorfer W.} \REVIEW{Nature Mater.}{7}{2008}{179}.


\bibitem{Mannini_NatureMater8/09}
\Name{Mannini M., Pineider F., Sainctavit P., Danieli C., Otero
E., Sciancalepore C., Talarico A.M., Arrio M.A., Cornia A.,
Gatteschi D. and Sessoli R.} \REVIEW{Nature Mater.}{8}{2009}{194}.




\bibitem{Timm_PRB73/06}
\Name{Timm C. \and Elste F.} \REVIEW{Phys. Rev. B}{73}{2006}{235304}.

\bibitem{Misiorny_PRB75/07}
\Name{Misiorny M. \and Barna{\'s} J.} \REVIEW{Phys. Rev. B}{75}{2007}{134425}.

\bibitem{Misiorny_PSSb246/09}
\Name{Misiorny M. \and Barna\'{s} J.} \REVIEW{Phys. Stat. Sol. B}{246}{2009}{695}.

\bibitem{Lu_PRB79/09}
\Name{Lu H.-Z., Zhou B. \and Shen S.-Q.} \REVIEW{Phys. Rev. B}{79}{2009}{174419}.

\bibitem{Barrazalopez_JAP105/09}
\Name{Barraza-Lopez S., Park K., Garc{\'\i}a-Su{\'a}rez V. \and Ferrer J.}
  \REVIEW{J. Appl. Phys.}{105}{2009}{07E309}.




\bibitem{Barrazalopez_PRL102/09}
\Name{Barraza-Lopez S., Park K., Garc{\'\i}a-Su{\'a}rez V. \and Ferrer J.}
  \REVIEW{Phys. Rev. Lett.}{102}{2009}{246801}.

\bibitem{Heersche_PRL96/06}
\Name{Heersche  H.~B., de Groot Z., Folk J.~A., van der Zant
H.~S.~J., Romeike C.,  Wegewijs  M.~R., Zobbi L., Barreca D.,
Tondello  E. \and Cornia, A.} \REVIEW{Phys. Rev.
Lett.}{96}{2006}{206801}.

\bibitem{Ni_APL89/06}
\Name{Ni  C., Shah, S., Hendrickson D. \and Bandaru, P.~R.} \REVIEW{Appl. Phys. Lett.}{89}{2006}{212104}.


\bibitem{Jo_NanoLett6/06}
\Name{Jo  M.-H., Grose J.~E., Baheti K., Deshmukh M.~M., Sokol
J.~J., Rumberger E.~M., Hendrickson D.~N., Long J.~R., Park H.
\and Ralph, D.~C.} \REVIEW {Nano Lett.}{6}{2006}{2014}.

\bibitem{Henderson_JApplPhys101/07}
\Name{Henderson J.~J., Ramsey C.~M., del Barco E., Mishra A., \and
Christou, G.} \REVIEW{J. Appl. Phys.}{101}{2007}{09E102}.

\bibitem{Voss_PRB78/08}
\Name{Voss  S., Zander O., Fonin M., R\"{u}diger U., Burgert M.
\and Groth, U.} \REVIEW{Phys. Rev. B}{78}{2008}{155403}.






\bibitem{Fonin_Polyhedron28/09}
\Name{Fonin M., Voss S., Herr S., de Loubens G., Kent A.D., Burgert M., Groth U., and R\"{u}diger U.}
  \REVIEW{Polyhedron}{28}{2009}{1977}.


\bibitem{Zobbi_ChemCommun12/05}
\Name{Zobbi L., Mannini M., Pacchioni M., Chastanet G., Bonacchi D., Zanardi C., Biagi R.,
Pennino U.D., Gatteschi D., Cornia A., and Sessoli R.}
  \REVIEW{Chem. Commun.}{12}{2005}{1640}.

\bibitem{Abdi_JAP95/04}
\Name{Abdi A.N., Bucher J.P., Rabu P., Toulemonde O., Drillon M., and Gerbier P.}
  \REVIEW{J. Appl. Phys.}{95}{2004}{7345}.


\bibitem{Cornia_StructBond122/05}
\Name{Cornia A., Costantino A.F., Zobbi L., Caneschi A., Gatteschi D., Mannini M., and Sessoli R.}
  \REVIEW{Struct. Bond.}{122}{2005}{133}.


\bibitem{Naitabdi_AdvMater17/05}
\Name{Naitabdi A., Bucher J.P., Gerbier P., Rabu P., and Drillon M.}
  \REVIEW{Adv. Mater.}{17}{2005}{1612}.


\bibitem{Burgert_JAmChemSoc129/07}
\Name{Burgert M., Voss S., Herr S., Fonin M., Groth U., and R\"{u}diger U.}
  \REVIEW{J. Am. Chem. Soc.}{129}{2007}{14362}.







\bibitem{Rudzinski_PRB64/01}
\Name{Rudzi{\'n}ski W. \and Barna{\'s} J.} \REVIEW{Phys. Rev. B}{64}{2001}{085318}.

\bibitem{Wilczynski_JMMM290/05}
\Name{Wilczy{\'n}ski M., {\'S}wirkowicz R., Rudzi{\'n}ski W., Barna{\'s} J.
  \and Dugaev V.} \REVIEW{J. Magn. Magn. Mater.}{290}{2005}{209}.

\bibitem{Souza_PRB75/07}
\Name{Souza F.~M., Egues J.~C. \and Jauho A.~P.} \REVIEW{Phys. Rev. B}{75}{2007}{165303}.

\bibitem{Hamaya_PRL102/09}
\Name{Hamaya K., Kitabatake M., Shibata K., Jung M., Ishida S., Taniyama T.,
  Hirakawa K., Y. A. \and Machida T.} \REVIEW{Phys. Rev. Lett.}{102}{2009}{236806}.

\bibitem{Merchant_PRL100/08}
\Name{Merchant C.~A. \and Markovi\'{c} N.} \REVIEW{Phys. Rev. Lett.}{100}{2008}{156601}.

\bibitem{Weymann_APL92/08}
\Name{Weymann I. \and Barna\'{s} J.} \REVIEW{Appl. Phys. Lett.}{92}{2008}{103127}.

\bibitem{Schoeller_PRB50/94}
\Name{Schoeller H. \and Sch{\"o}n G.} \REVIEW{Phys. Rev. B}{50}{1994}{18436}.

\bibitem{Koenig_PRB54/96}
\Name{K{\"o}nig J., Schmid J., Schoeller H. \and Sch{\"o}n G.} \REVIEW{Phys.
  Rev. B}{54}{1996}{16820}.

\bibitem{Misiorny_PRB79/09}
\Name{Misiorny M., Weymann I. \and Barna{\'s} J.} \REVIEW{Phys. Rev. B}{79}{2009}{224420}.

\bibitem{Blanter_PhysRep336/00}
\Name{Blanter Y. \and B\"{u}ttiker M.} \REVIEW{Phys. Rep.}{336}{2000}{1}.

\bibitem{Thielmann_PRB68/03}
\Name{Thielmann A., Hettler M., K{\"o}nig J. \and Sch{\"o}n G.} \REVIEW{Phys.
  Rev. B}{68}{2003}{115105}.

\bibitem{Weymann_PRB78/08}
\Name{Weymann I.} \REVIEW{Phys. Rev. B}{78}{2008}{045310}.

\end{thebibliography}
\end{document}